\documentclass[aps,prl,floatfix,epsfig,twocolumn,showpacs,preprintnumbers]{revtex4}
\usepackage{amsmath}
\usepackage{graphicx}
\usepackage{epstopdf}
\usepackage{dcolumn}
\usepackage{bm}
\usepackage{mathrsfs}
\usepackage[pdfstartview=FitH, CJKbookmarks=true, bookmarksnumbered=true, bookmarksopen=true, pdfborderstyle={/S/U}, pdfborder={0 0 1}, colorlinks=true, linkcolor=blue, anchorcolor=green, citecolor=blue]{hyperref}

\begin{document}

\title{A Monopole Mining Method for High Throughput Screening Weyl
Semimetals }
\author{Vsevolod Ivanov and Sergey Y. Savrasov}
\affiliation{Department of Physics, University of California, Davis, CA 95616, USA}

\begin{abstract}
Although topological invariants have been introduced to classify the
appearance of protected electronic states at surfaces of insulators, there
are no corresponding indexes for Weyl semimetals whose nodal points may
appear randomly in the bulk Brillouin Zone (BZ). Here we use a well--known
result that every Weyl point acts as a Dirac monopole and generates integer
Berry flux to search for the monopoles on rectangular BZ\ grids that are
commonly employed in self--consistent electronic structure calculations. The
method resembles data mining technology of computer science and is
demonstrated on locating the Weyl points in known Weyl semimetals. It is
subsequently used in high throughput screening several hundreds of compounds
and predicting a dozen new materials hosting nodal Weyl points and/or lines.
\end{abstract}

\date{\today }
\maketitle

There has been recent surge of interest in topological quantum materials
caused by the existence in these systems of robust electronic states
insensitive to perturbations\cite{RMPTI, RMPWSM}. Z$_{2}$ invariants have
been proposed to detect the protected (quantum Hall--like) surface states in
topological insulators (TIs) \cite{Z2}, and, for centrosymmetric crystals,
this reduces\ to finding band parities of electronic wave functions at
time--reversal invariant points in the Brillouin zone (BZ)\cite{FuKane}. For
a general case, the calculation involves an integration of Berry fields \cite%
{Berry}, and has been implemented in numerical electronic structure
calculations\cite{nfield} with density functional theory. These methods have
allowed for exhaustive searches to identify candidate materials hosting
topological insulator phases \cite{Hasan,Zhang,Bernevig}.

Weyl semimetals (WSMs) are closely related systems characterized by a bulk
band structure which is fully gapped except at isolated points described by
the 2x2 Weyl Hamiltonian \cite{RMPWSM}. Sometimes these Weyl points extend
into lines in the BZ giving rise to nodal line semimetals (NLSMs) \cite{NLM}%
. Due to their intriguing properties such as Fermi arc surface states \cite%
{Arcs}, chiral anomaly induced negative magnetoresistance \cite{Nielsen},
and a semi--quantized anomalous Hall effect \cite{Ran,Balents}, the search
for new WSM materials is currently very active. Unfortunately, their
identification in infinite space of chemically allowed compounds represents
a challenge: there is no corresponding topological index characterizing WSM
phase, and the Weyl points may appear randomly in the bulk BZ. General
principles, such as broken time reversal or inversion symmetry, or emergence
of the WSM phase between topologically trivial and non--trivial insulating
phases \cite{Arcs} are too vague to guide their high throughput screening,
and recent\ group theoretical arguments\cite{Po,Krutho} to connect crystal
symmetry with topological properties still await their practical
realization. The progress in this field was mainly serendipitous, although
the ideas based on\ band inversion mechanism\cite{BHZ} or analyzing mirror
Chern numbers\cite{MCN1,MCN2} were proven to be useful in many recent
discoveries\cite{TaAs,LaAlGe,Ta3S2,CuF}, and computer oriented searches of
topological semimetals are beginning to appear \cite{Z2pack,more}. 
\begin{figure}[tbp]
\includegraphics[height=0.20\textwidth,width=0.40\textwidth]{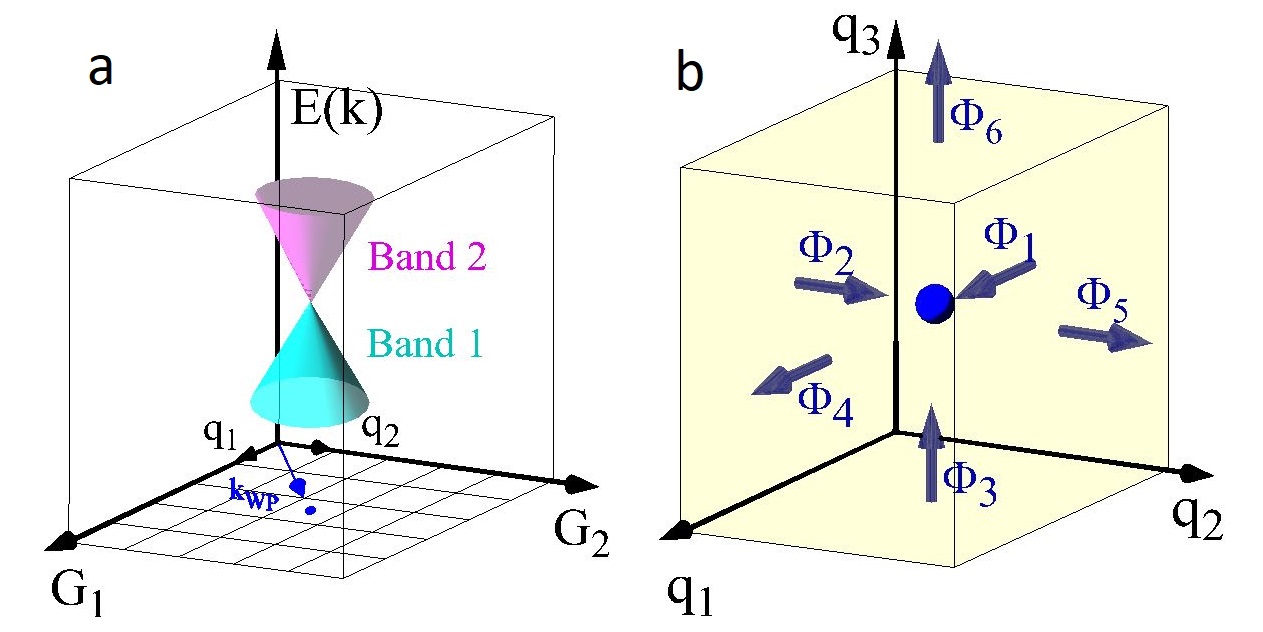}
\caption{a. A typical cone dispersion relationship $E(\bf{k})$=$\pm v|%
\bf{k}-\bf{k}_{WP}|$ for the Weyl point plotted within a rectangular
area in k--space set by divisions of reciprocal lattice translations $%
\bf{G}_{1}$ and $\bf{G}_{2}$ for a fixed value along the third
translation $\bf{G}_{3}$. b. The Weyl point located within a microcell
set by the grid vectors $\bf{q}_{1},\bf{q}_{2},\bf{q}_{3}$
generates a Berry flux through each plaquette as given by the (right handed)
circulation of the Berry connection with sign convention defined in text. }
\label{FigMethod}
\end{figure}

In this work, we propose a straightforward method to identify Weyl
semimetals by using a well--known result that every Weyl point acts as a
Dirac monopole \cite{Fang} producing a non--zero Berry flux when it is
completely enclosed by a surface in the BZ. The enclosed charge is given by
the chirality of the Weyl point similar to the Gauss theorem in the Coulomb
law. Rectangular grids of k--points that are widely employed in
self--consistent electronic structure calculations for the BZ integration
either via special points (Monkhorst-Pack) technique \cite{MP} or a
tetrahedron method\cite{Tetra}, are ideally suited for this purpose since
they divide the volume of the BZ onto microcells and the electronic wave
functions are automatically available at the corners of each microcell. It
is thus a matter of rearranging the data to extract Berry phases of these
wave functions in order to recover the Dirac monopoles inside the BZ. While
there are some uncertainties connected to energy bands cutoffs used while
defining non--Abelian Berry fields for metallic systems, our method allows a
subsequent refinement provided a signal from a monopole is detected. The
entire procedure resembles data mining technology in computer science as an
intelligent method to discover patterns from large data sets in a (semi--)
automatic way so that the extracted data can subsequently be used in further
analysis.

Since we are dealing with grids, there is a chance that the grid microcell
will enclose both chiral positive and negative charges whose Berry fluxes
cancel each other. Although resolution here is obviously adjustable by
changing the grid size, and modern computers allow handlings of thousands
and even millions of k--points in parallel, going for Weyl points that are
too close makes no sense from both practical and fundamental reasons.
Practically, properties such as anomalous Hall effect\cite{Ran,Balents} are
proportional to the distance between the Weyl points and so does the density
of Fermi arc surface states\cite{Arcs}. Disorder, electronic interactions,
thermal broadening and Heisenberg uncertainty principle provide fundamental
limitations. Therefore, distances between the Weyl points need not be
smaller than a few percent of the reciprocal lattice spacing, and this does
not require dealing with very dense grids.

\begin{figure}[tbp]
\includegraphics[height=0.27\textwidth,width=0.40\textwidth]{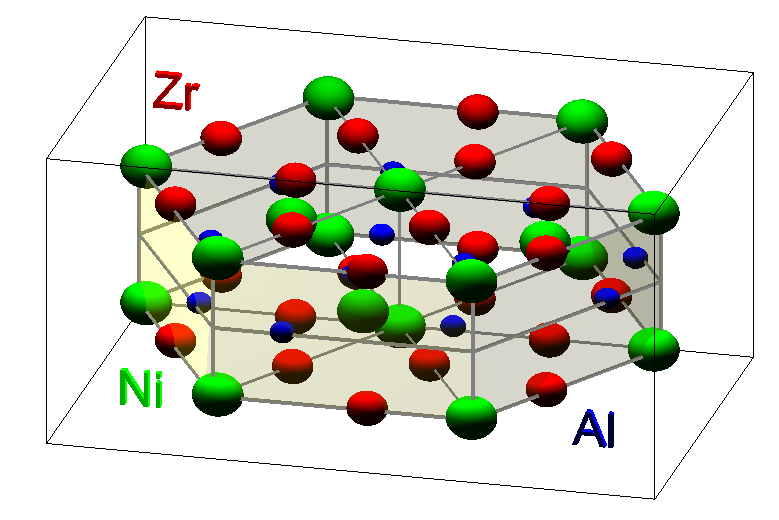}
\caption{ZrNiAl--type crystal structure (\# 189 space group $p\bar{6}2m$) of
noncentrosymmetric hexagonal compounds compounds studied in this work.}
\label{FigStruc}
\end{figure}

Here we implement this monopole mining method and test it by verifying
locations of the Weyl points in several known systems, such as recently
proposed TaAs\cite{TaAs} and CuF\cite{CuF} Weyl semimetals. Next, we
demonstrate how it can be used for high throughput screening of WSMs by
scanning several hundreds of compounds in the $p\bar{6}2m$(\#189) space
group with the ZrNiAl structure. We predict a dozen new materials hosting
WSM/NLSM behavior.

We first outline the method to evaluate the Berry flux due to a single Weyl
point that appears somewhere in the bulk BZ with its typical dispersion
relationship $E(\bf{k})=\pm v|\bf{k}-\bf{k}_{WP}|$ as
illustrated in Fig.\ref{FigMethod}a. We represent the BZ by reciprocal
lattice translations $\bf{G}_{\nu =1,2,3}$ and divide it onto $%
N_{1}\times N_{2}\times N_{3}$ microcells. Each microcell is spanned by
primitive vectors $\bf{q}_{\nu =1,23}=\bf{G}_{\nu }/N_{\nu }$ with
its origin given by the grid of $\bf{k}$--points represented by three
integers $n_{\nu }=0,N_{\nu }-1$ as $\bf{k}=n_{1}\bf{q}_{1}+n_{2}%
\bf{q}_{2}+n_{3}\bf{q}_{3}.$

The problem of finding the wave vector $\bf{k}_{WP}$ is reduced to
recovering the microcell that contains the monopole. We define a
non--Abelian link field that appears while evaluating the Berry phase using
the finite difference method\cite{nfield}%
\begin{equation}
U_{\bf{q}}(\bf{k})=\frac{\det \left[ \langle \bf{k}+\bf{q}%
j^{\prime }|e^{i\bf{qr}}|\bf{k}j\rangle \right] }{\left\vert \det %
\left[ \langle \bf{k}+\bf{q}j^{\prime }|e^{i\bf{qr}}|\bf{k}%
j\rangle \right] \right\vert }  \label{ULINK}
\end{equation}%
Here the matrix elements between the periodic parts of the wave functions
are cast into the form $\langle \bf{k}+\bf{q}j^{\prime }|e^{i\bf{%
qr}}|\bf{k}j\rangle ,$ which frequently appear in density functional
linear response calculations\cite{LRT} and thus are straightforward to
evaluate. The set of energy bands $j\ $is spanned over occupied states and
includes those that cross the Fermi level. However, some uncertainty exists
in this enumeration procedure because the Berry flux from the negative and
positive branches\ of the monopole (bands 1 and 2 for the example shown in
Fig.\ref{FigMethod}a)will cancel each other. For the example being
discussed, this means that either band 1 or 2 (but not both) needs to be
taken into account while evaluating Eq.\ref{ULINK}. In real materials, this
may result in contribution for some monopoles cancelling, but since we are
mostly interested in the Weyl points in the immediate vicinity of the Fermi
level, varying the upper cutoff value for $j$ by one or two will resolve
this problem. We also note that the link field $U_{\bf{q}}(\bf{k})$
needs to be computed for the entire grid of k--points, where the group
symmetry operations help to generate the wave functions that are normally
available within only irreducible portion of the BZ.

We now evaluate the Berry flux through faces of each microcell of the $%
N_{1}\times N_{2}\times N_{3}$ grid. This is illustrated in Fig.\ref%
{FigMethod}b, where the flux $\Phi _{i=1..6}\ $through\ each plaquette with
the origin at particular $\bf{k}$ and spanned by a pair of vectors $%
q_{\mu }q_{\nu }$ is conveniently encoded into the following formula 
\begin{equation}
2\pi \Phi \equiv Im\ln \left[ \frac{U_{\bf{q}_{\mu }}(\bf{k}%
)U_{\bf{q}_{\nu }}(\bf{k+q}_{\mu })}{U_{\bf{q}_{\nu }}(\bf{k}%
)U_{\bf{q}_{\mu }}(\bf{k+q}_{\nu })}\right]  \label{FLUX}
\end{equation}%
This procedure is similar to one employed while evaluating $Z_{2}$
invariants \cite{nfield} on six two--dimensional tori introduced in Ref. 
\cite{Moore} but now the roles of the tori are played by the slices of the
BZ spanned by each pair of the reciprocal vectors $G_{\mu }G_{\nu }$ with a
fixed value along the third vector $G_{\xi }$. We only need to take care of
the fact that the flux as given by Eq. \ref{FLUX} produces right
(alternatively left) handed circulation of the Berry connection but inner
(or outer) normal should be chosen consistently for the total flux through
each surface of the microcell. Thus, the total Berry flux is given by%
\begin{equation}
c=\Phi _{1}+\Phi _{2}+\Phi _{3}-\Phi _{4}-\Phi _{5}-\Phi _{6}  \label{CHARGE}
\end{equation}%
Although the flux through each plaquette is generally non--integer, the
total flux is guaranteed to be an integer since individual contributions (%
\ref{FLUX}) from adjacent plaquettes cancel each other in Eq.(\ref{CHARGE}),
up to an addition of $2\pi n$. Therefore$\ c$ returns ether the chiral
charge of the monopole or zero.

The entire algorithm is now viewed as an automated procedure that is either
done following the self--consistent band structure calculation or "on the
fly". We illustrate it on the example of TaAs Weyl semimetal whose
electronic properties are well documented in recent literature \cite{TaAs}.
We use a full potential linear muffin--tin orbital method (FP LMTO)
developed by one of us \cite{FPLMTO} and perform a self--consistent density
functional calculation with spin--orbit coupling using the Generalized
Gradient Approximation \cite{GGA}. We subsequently set up a k--grid using $%
20\times 20\times 20$ divisions of the reciprocal lattice unit cell. These
types of grids were previously shown to be sufficient in calculating Z$_{2}$
invariants in topological insulators\cite{CPC}. For evaluating the link
field, Eq. (\ref{ULINK}), the energy window is chosen to span the entire
valence band with the cutoff value corresponding to the band number that
crosses the Fermi level. It appears this is sufficient to recover all
monopoles. The net result is 24 out 8000 microcells produce non--zero Berry
flux and give their approximate positions. We take the coordinates of the
corresponding microcells (only non--equivalent by symmetry are needed; two
for TaAs) and mine these areas of k--space by introducing similar
rectangular grids inside each microcell in order to refine the locations of
the Weyl points to the positions: $(0.009,0.506,0)$, $(0.019,0.281,0.579)$
in units $2\pi /a,2\pi /a,2\pi /c.$ This is in agreement with the previous
calculation \cite{TaAs}.

We also considered CuF, recently predicted to be a Weyl semimetal by one of
us\cite{CuF}. The exact same setup ($20\times 20\times 20$ divisions with
the energy panel spanned till the band that crosses the Fermi level) returns
24 microcells that are all related by symmetry. Zooming into one microcell
returns the following location of the Weyl point: $(0.281,0.119,0)2\pi /a,$
consistent with our previous result \cite{CuF}.

To demonstrate the predictive power of the method, we scanned several
hundreds noncentrosymmetric hexagonal compounds in the $p\bar{6}2m$ (\# 189)
space group with the ZrNiAl structure. A complete list of these materials is
given in Supplementary Infortmation. Topological electronic structures in
few of these systems have already drawn a recent attention. CaAgP was
predicted to be a line--node Dirac semimetal while CaAgAs was found to be a
strong topological insulator \cite{CaAgAs}. Similar properties have been
discussed for NaBaBi under pressure\cite{NaBaBi}. The unit cell of these
crystals consists of a rhomboid prism with side $a$, internal angle $2\pi /3$%
, and height $c$; the Ni-type atoms are located on the vertical edges and in
the centers of the two equilateral triangles forming the rhombus base, with
the Zr-type and Al-type atoms located on the edges of these triangles, $%
1/3\pm 1/4(c^{2}/a^{2})$ away from the corners in the middle and bottom
layers respectively [see Fig. \ref{FigStruc}]. We perform self--consistent
band structure calculations and subsequent monopole mining procedure in
exactly the same manner as we illustrated for TaAs and CuF. The lattice
parameters can be found in Ref. \cite{Struc}.

\begin{table}[tbp]
\caption{List of non--equivalent Weyl and triple points (in units $2\protect%
\pi /a,2\protect\pi /a,2\protect\pi /c)$, their number and energies relative
to the Fermi level (in eV) recovered using the monopole mining method for \
noncentrosymmetric hexagonal compounds in the $p\bar{6}2m$ (\# 189) space
group with the ZrNiAl structure that are predicted to exhibit Weyl/nodal
line semimetal behavior. The typical appearance of the Weyl points in the
Brillouin Zone is cited by referencing to either sort A or B as illustrated
in Fig. \protect\ref{Weyls}ab.}
\label{TablePoints}%
\begin{tabular}{|l|l|l|l|l|}
\hline\hline
Comp. & Topological Points & Type & \# & E (eV) \\ \hline
LaInMg & $(0.00000,0.36868,0.01123)$ & Weyl-A & $12$ & $%
\begin{array}{c}
-0.06%
\end{array}%
$ \\ 
LuGeAg & $(0.00000,0.42190,0.00098)$ & Weyl-A & $12$ & $%
\begin{array}{c}
-0.23%
\end{array}%
$ \\ 
YGeLi & $(0.00000,0.27793,0.00817)$ & Weyl-A & $12$ & $%
\begin{array}{c}
-0.13%
\end{array}%
$ \\ 
YPbAg & $(0.00000,0.40335,0.03142)$ & Weyl-A & $12$ & $%
\begin{array}{c}
-0.09%
\end{array}%
$ \\ 
YSiAg & $(0.00000,0.37864,0.00384)$ & Weyl-A & $12$ & $-0.09$ \\ 
HfPRu & $(0.46280,0.06931,0.0221)$ & Weyl-B & $24$ & $+0.06$ \\ 
ZrPRu & $(0.45982,0.07532,0.01698)$ & Weyl-B & $24$ & $+0.06$ \\ \hline
LaTlMg & $%
\begin{array}{c}
(0.00000,0.38916,0.03236) \\ 
(0.41450,0.02567,0.00724)%
\end{array}%
$ & $%
\begin{array}{c}
\text{Weyl-A} \\ 
\text{Weyl-B}%
\end{array}%
$ & $%
\begin{array}{c}
12 \\ 
24%
\end{array}%
$ & $%
\begin{array}{c}
-0.13 \\ 
-0.13%
\end{array}%
$ \\ \hline
YTlMg & $%
\begin{array}{c}
(0.00000,0.43303,0.02319) \\ 
(0.44076,0.02908,0.00441)%
\end{array}%
$ & $%
\begin{array}{c}
\text{Weyl-A} \\ 
\text{Weyl-B}%
\end{array}%
$ & $%
\begin{array}{c}
12 \\ 
24%
\end{array}%
$ & $%
\begin{array}{c}
-0.05 \\ 
-0.11%
\end{array}%
$ \\ \hline
LuAsPd & $%
\begin{array}{c}
(0.00000,0.11481,0.14140) \\ 
(0.00000,0.12004,0.13942)%
\end{array}%
$ & $%
\begin{array}{c}
\text{Weyl-A} \\ 
\text{Weyl-A}%
\end{array}%
$ & $%
\begin{array}{c}
12 \\ 
12%
\end{array}%
$ & $%
\begin{array}{c}
+0.18 \\ 
+0.19%
\end{array}%
$ \\ \hline
ZrAsOs & $%
\begin{array}{c}
(0.47365,0.02591,0.04792) \\ 
0.47406,0.01215,0.04789%
\end{array}%
$ & $%
\begin{array}{c}
\text{Weyl-B} \\ 
\text{Weyl-B}%
\end{array}%
$ & $%
\begin{array}{c}
24 \\ 
24%
\end{array}%
$ & $%
\begin{array}{c}
+0.02 \\ 
+0.02%
\end{array}%
$ \\ \hline
TiGePd & $%
\begin{array}{c}
(0.00000,0.00000,0.16495) \\ 
(0.00000,0.00000,0.20775)%
\end{array}%
$ & $%
\begin{array}{c}
\text{Triple} \\ 
\text{Triple}%
\end{array}%
$ & $%
\begin{array}{c}
2 \\ 
2%
\end{array}%
$ & $%
\begin{array}{c}
+0.14 \\ 
+0.22%
\end{array}%
$ \\ \hline
VAsFe & $%
\begin{array}{c}
(0.00000,0.000000,0.32279) \\ 
(0.00000,0.000000,0.47625) \\ 
(0.00000,0.38339,0.17269)%
\end{array}%
$ & $%
\begin{array}{c}
\text{Triple} \\ 
\text{Triple} \\ 
\text{Weyl-A}%
\end{array}%
$ & $%
\begin{array}{c}
\text{2} \\ 
\text{2} \\ 
1\text{2}%
\end{array}%
$ & $%
\begin{array}{c}
+0.14 \\ 
+0.19 \\ 
+0.09%
\end{array}%
$ \\ \hline
\end{tabular}%
\end{table}

Out of the compounds that we studied, we clearly identify 11 materials which
show WSM\ behavior, 1 NLSM and 1 hosting both Weyl points and nodal lines.
The two NLSMs also host topologically distinct triple fermion points \cite%
{TP}. Table \ref{TablePoints} summarizes our results for each compound ,
giving the locations of the non--equvalent low--energy Weyl and/or triple
points, their number and energies relative E$_{F}$ in eV. The Weyl points
are generally viewed as type II\ according to classification introduced in
Ref. \cite{TypeII}. (Complete crystallographic and electronic structure data
for these compounds is given in the supplementary information.)

\begin{figure*}[tbh]
\center\includegraphics[height=0.285\textwidth,width=0.95%
\textwidth]{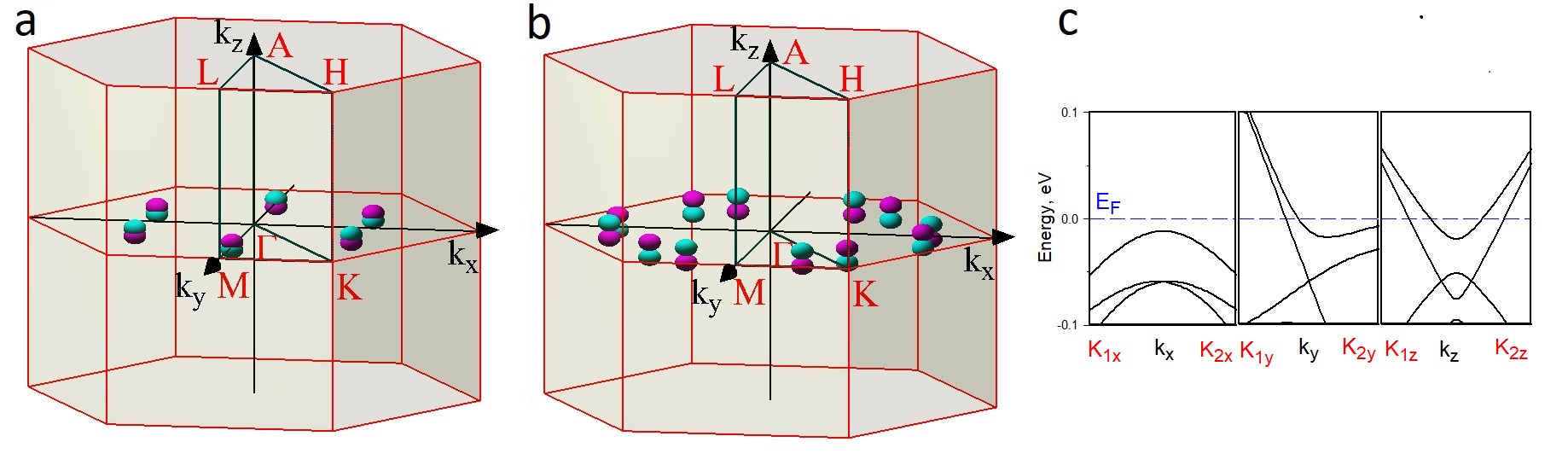}
\caption{a. Positions of 6 pairs (cyan for chiral positive and magenta for
chiral negative) of low--energy Weyl points seen along the $\Gamma M$
direction in the BZ for LaInMg and referenced in Table \protect\ref%
{TablePoints} as sort A; b. Positions of 12 pairs of Weyl points that are
shifted symmetrically away from the $\Gamma K$ line for HfPRu and referenced
in Table \protect\ref{TablePoints} as sort B; c. Energy band dispersions in
the vicinity of the Weyl point $k_{wp}=(0.00000,0.36868,0.01123)$ for
LaInMg. Point notations are as follows: $%
K_{1x}=(-0.10000,0.36868,0.01123),K_{2x}=(0.10000,0.36868,0.01123),K_{1y}=(0.0000,0.26868,0.01123),K_{2y}=(0.0000,0.46868,0.011230),K_{1z}=(0.0000,0.36868,-0.056150),K_{2z}=(0.0000,0.36868,0.056150) 
$ in units $2\protect\pi /a,2\protect\pi /a,2\protect\pi /c$. }
\label{Weyls}
\end{figure*}

Many of the Weyl semimetals that we predict in our work display remarkably
similar locations of their Weyl points. LaInMg, LuGeAg, YGeLi,YPbAg, and
YSiAg, exhibit 6 pairs (chiral positive and negative) of points, that are
all symmetry related and are only slightly displaced from the k$_{z}=0$
plane. They are located along the $\Gamma M$ direction in the BZ. We
illustrate their precise positions for LaInMg in Fig. \ref{Weyls}a and refer
to them in Table \ref{TablePoints} as Weyl points of sort A. We find that
HfPRu, and ZrPRu show another sort (referred to as sort B) of Weyl points,
namely 12 pairs that are shifted symmetrically away from the $\Gamma K$ line
(see Fig. \ref{Weyls}b). Interestingly, a similar behavior is seen for
LaTlMg, and YTlMg which show both sorts (A and B) of Weyl points. LuAsPd
shows two kinds of sort A Weyl points (24 total), while$\ $ZrAsOs shows two
kinds of sort B Weyl points (48 total). Their displacement from k$_{z}=0$
plane is much larger than the one found in previous cases. For each reported
Weyl point, we also provide independent verification by calculating the band
structures along $k_{x},k_{y}$ and $k_{z}$ directions with the boundary
vectors confining the Weyl point. An example of such plot is shown in Fig. %
\ref{Weyls}c for the Weyl point in LaInMg, where one clearly recognizes the
band crossings along all three directions that are characteristic of the
Weyl cone dispersion.

\begin{figure}[tbp]
\includegraphics[height=0.742\textwidth,width=0.40\textwidth]{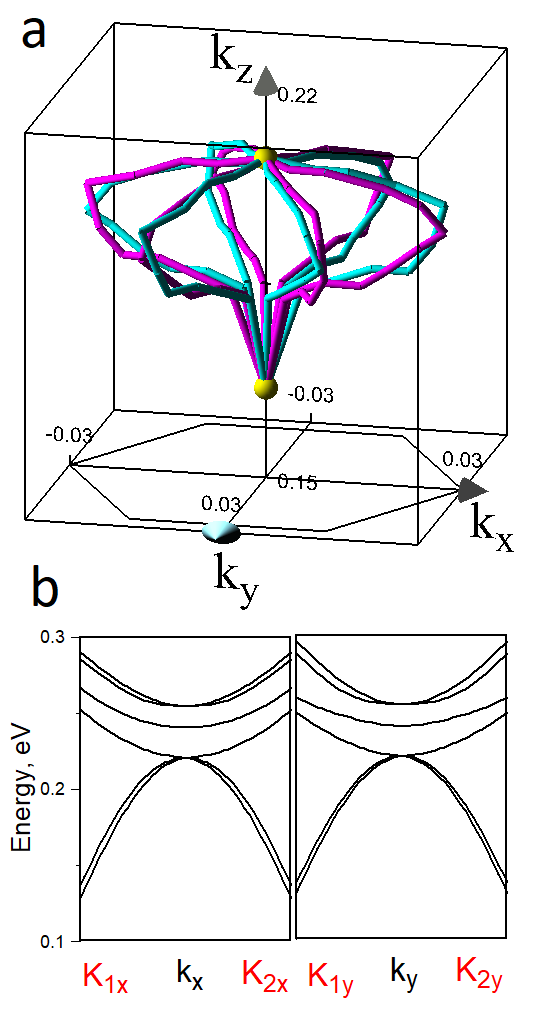}
\caption{a. A set of nodal lines for TiGePd that is recovered by the
monopole mining method presented in this work. The color (cyan and magenta)
distinguishes chiral positive and negative lines, respectively. The zoomed
area of the BZ is bounded by $0.15\leq 2\protect\pi k_{z}/c\leq 0.22$ and $%
-0.03\leq 2\protect\pi k_{x,y}/a\leq +0.03.$ Also shown in yellow are the
triple degenerate topological points \protect\cite{TP}. b. Energy band
dispersions in the vicinity of the triple point $(0,0,0.20775)$ for TiGePd.
Point notations are as follows: $%
K_{1x}=(-0.10000,0.0000,0.20775),K_{2x}=(0.10000,0.0000,0.20775),K_{1y}=(0.0000,-0.10000,0.20775) 
$ in units $2\protect\pi /a,2\protect\pi /a,2\protect\pi /c.$}
\label{Lines}
\end{figure}

Another interesting outcome of our high--throughput screening is the
materials exhibiting nodal lines and triple--point fermions. TiGePd and
VAsFe both host 12 pairs (chiral positive and negative) of nodal lines that
are located very close to the $\Gamma A$ direction in the BZ. We illustrate
this behavior for TiGePd in Fig. \ref{Lines}a by zooming into the area of
the BZ bounded by $0.15\leq 2\pi k_{z}/c\leq 0.22$ and $-0.03\leq 2\pi
k_{x,y}/a\leq +0.03.$ Interestingly, the nodal lines start and end at triple
degenerate points that have recently enriched our classification of the
topological objects \cite{TP}. These triple points are located at the $%
\Gamma A$ line of the BZ. We provide their coordinates for TiGePd and VAsFe
in Table \ref{TablePoints}. The corresponding band structure plot for one of
the triple points in TiGePd is shown in Fig. \ref{Lines}b along the $k_{x}$
and $k_{y}$ directions of the BZ with the boundary vectors confining the
triple point. (A complete set of plots for each compound is provided in the
supplementary information.)

One of the most striking features of Weyl semimetals is the presence of the
Fermi arcs in their one--electron surface spectra\cite{Arcs}. Although
computations of their shapes are possible via a self--consistent supercell
(slab) calculation of the surface energy bands, given the number of
compounds that we deal in this work, it is a computationally demanding
study. Nevertheless, since the arcs connect the Weyl points of different
chirality, one can expect that most of the materials that we list in Table %
\ref{TablePoints} would have rather short arcs since the distances between
positive and negative chiral charges are quite small. One notable exception
is VAsFe which, as we list in Table \ref{TablePoints}, exhibits not only
nodal lines and triple points, but also a set of Weyl points which are well
separated from each other. These are expected to produce very long Fermi
arcs for the (100) or (110) crystallographic types of surfaces. One can also
expect that their contribution to the anomalous Hall coefficient should be
large since the latter is known to be directly proportional to the distance
between the Weyl points \cite{Ran}. We have recently shown \cite{Disorder}
that long and straight Fermi arcs are generally capable of supporting nearly
dissipationless surface currents, therefore it could be interesting to
explore such possibility in VAsFe.

In conclusion, using the well--known property that Weyl points act as Dirac
monopoles in k--space, we presented an automated monopole mining method to
identify Weyl and nodal line semimetals. We tested the method by recovering
the Weyl points in several known systems as well as demonstrating its
predictive power by high throughput screening hundreds noncentrosymmetric
hexagonal compounds in the $p\bar{6}2m$ (\# 189) space group and finding 13
new materials whose electronic structures as well as the locations of the
topological nodal points and lines have been reported. As we judge from our
calculated energy bands, the WSMs identified in this work exhibit regular
Fermi surface states while the Weyl points are not exactly pinned at the
Fermi level. This is similar to other recently discovered WSMs, such as TaAs%
\cite{TaAs} whose experimental studies of large negative magnetoresistance
have been recently performed\cite{NegativeMR}. Despite the latter
representing a signature of the much celebrated chiral anomaly feature in
Weyl semimetals, there exists an obvious problem of distinguishing
contributions from the Weyl points and regular Fermi states. In this regard
our automated approach should be helpful for scanning vast material
databases in identifying an ideal WSM with only nodal points at the Fermi
level as it was originally envisioned in pyrochlore iridates\cite{Arcs}.

The work was supported by NSF DMR Grant No. 1411336.

\end{document}


\section{\textbf{SUPPLEMENTARY INFORMATION}}

\subsection{List of Compounds}

Here we list noncentrosymmetric hexagonal compounds in the $p\bar{6}2m$ (\#
189) space group with the ZrNiAl structure studied in this work. Their
complete crystallographic data can be found in Ref. \cite{Struc}. As many of
the compounds in this structure include rare earth elements with their f
electron states appearing in the vicinity of the Fermi level, we first
provide a list of only those compounds that do not explicitly include
Lanthanides (see Table \ref{TableCompounds}). These are the systems for
which density functional based calculations can be trusted in general.

We can also comment on the compounds that include Lanthanide elements. They
can be separated onto two large groups. The first group includes the
materials where the narrow f--band appears crossing the Fermi level in the
calculated band structures. This would be an indication that a many--body
renormalization of the single particle spectra (such, e.g., as band
narrowing, multiplet transitions, etc) is expected. Although modern
electronic structure approaches based on combinations of density functional
and dynamical mean field theories \cite{DMFT-RMP} allow handling such cases,
those are outside the scope of the present study, and we do not study
topological properties of these compounds. The second group includes the
materials with either fully empty or fully occupied f band, namely f$^{0}:$
LaAuCd, LaAuIn, LaAuMg, LaCuIn, LaCuMg, LaInMg, LaIrSn, LaNiIn, LaNiZn,
LaPdCd, LaPdHg, LaPdIn, LaPdMg, LaPdPb, LaPdSn, LaPdTl, LaPtIn, LaPtPb,
LaPtSn, LaRhIn, LaRhSn, LaTlMg; f$^{14}$: LuAsPd, LuAuIn, LuAuZn, LuCuIn,
LuGaMg LuGeAg, LuGeLi, LuInMg, LuIrSn, LuNiAl LuNiIn, LuNiPb, LuPbAg,
LuPdIn, LuPdSn LuPdZn, LuPtIn, LuPtSn, LuRhSn, LuSiAg, LuTlMg. These are the
cases where static mean field description can in principle capture single
particle excitations (apart from the question whether the position of the
f--band is correctly predicted by such theory).

\begin{table}[tbp]
\caption{List of noncentrosymmetric hexagonal compounds in the $p\bar{6}2m$
(\# 189) space group with the ZrNiAl structure\ studied in this work. The
compounds containing Lanthanide element are explicitly excluded from the
Table. }
\label{TableCompounds}%
\begin{tabular}{llll}
\hline
Class & X = & Class & X = \\ \hline
CrAsX & Ti, Pd, Fe, Co, Ni, Rh & XPtIn & Sc, Y \\ 
MnAsX & Ti, Ni, Rh, Fe, Pd, Ru & TiGeX & Co, Pd \\ 
ScGeX & Fe, Rh, Cu, Os, Pd, Ru & ZrCoX & Ga, Sn \\ 
XSiRe & Hf, Ta, Ti, Zr & ZrGeX & Os, Zn \\ 
HfGeX & Fe, Os, Rh, Ru & XNiGa & Hf, Zr \\ 
FeAsX & Ti, Co, V , Ni & ScPX & Ir, Na \\ 
XPNi & Fe, Mo, W , Co & MnGeX & Pd, Rh \\ 
XGeMn & Hf, Nb, Sc, Ta & CrPX & Pd, Ni \\ 
TiPX & Cr, Os, Ru & HfXRu & P, As \\ 
ZrPX & Os, Mo, Ru & XAsOs & Hf, Zr \\ 
MnPX & Rh, Pd, Ni & XPdPb & Ca, Y \\ 
ScSiX & Cu, Ru, Mn & XSiMn & Nb, Ta \\ 
CaXCd & Ge, Sn, Pb & HfSiX & Os, Ru \\ 
XAsPd & Hf, Ti, Zr & CaXAg & P, As \\ 
XNiAl & Hf, Y , Zr & ZrXRu & Si, As \\ 
XBFe & Nb, Ta & NbCrX & Ge, Si \\ \hline
\multicolumn{4}{l}{Other: YRhSn, YAuCd , YPdMg , YNiIn , ScGeAg} \\ 
\multicolumn{4}{l}{YPdAl , YInMg , YAuZn , YPbAg , YPdTl} \\ 
\multicolumn{4}{l}{YAuMg , YPdZn , YPdTl , YSiAg , YRhIn} \\ 
\multicolumn{4}{l}{YTlMg , YAgMg , YAuIn , YCuIn , YGaMg} \\ 
\multicolumn{4}{l}{HfIrSn , YPdIn , YCuAl , YGeLi , YPtSn} \\ 
\multicolumn{4}{l}{YCuMg , BaBiNa, YAlMg, ScSnAg, YSiLi} \\ \hline
\end{tabular}%
\end{table}

There are a few materials that include Sm ion with its non--magnetic
configuration f$^{6}:$ SmAgMg, SmAuCd, SmAuIn, SmAuMg, SmCuAl, SmCuIn,
SmIrIn, SmIrSn, SmNiAl, SmNiIn, SmNiSn, SmNiZn, SmPdCd, SmPdHg, SmPdIn,
SmPdMg, SmPdPb, SmPdTl, SmPtIn, SmPtMg, SmPtPb, SmPtSn, SmRhIn, SmRhSn,
SmSiAg, SmTlMg. Here $j=5/2$ and $j=7/2$ subbands appear below and above the
Fermi level, respectively. The Coulomb renormalzation in these compounds has
a predictable effect by renormalizing the spin--orbit coupling by the
Hubard--type interaction, and the states in the immidiate vicinity of the
Fermi level are not affected.

\subsection{Data for Topological Points}

Figures 1--13 provide complete data for for the topological materials
predicted in this work: the band structures near the Fermi level, energy
panels used for defining non--Abelian Berry connection, positions of
low--energy topological nodal points in the Brillouin Zone as well as energy
band dispersions in the vicinity of the nodal points.

\begin{widetext}

\begin{center}
\begin{figure}[h]
\includegraphics[height=0.304\textwidth,width=0.95\textwidth]{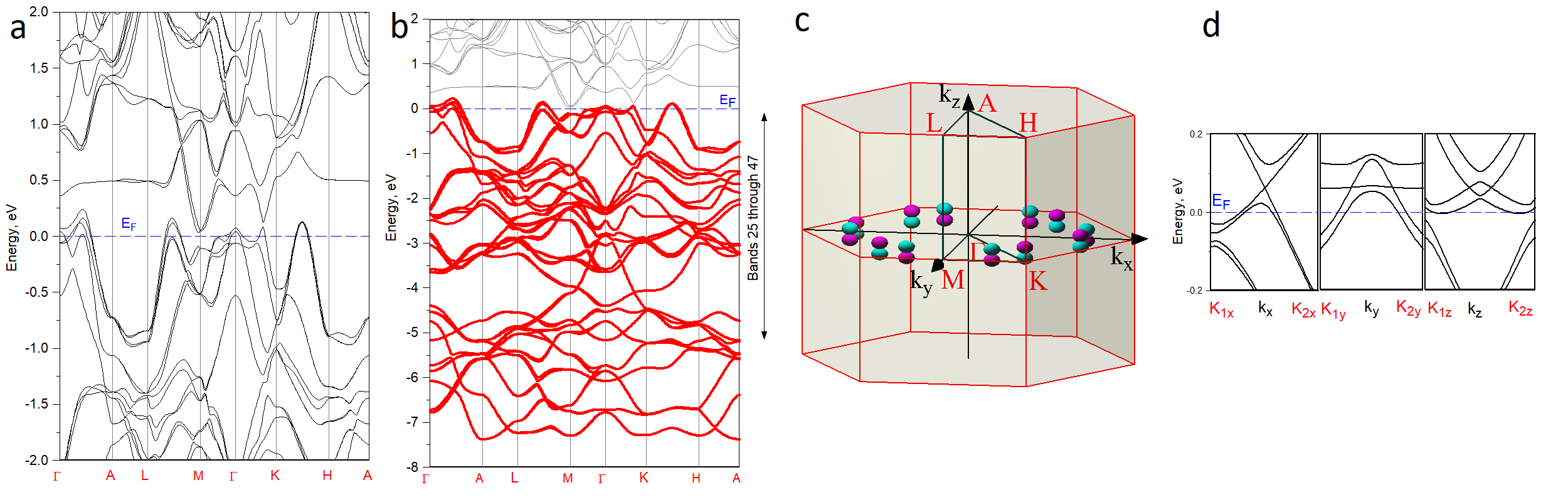}
\caption{Results for HfPRu: a. band structure near the Fermi level; b.
energy panel used for defining non-Abelian Berry connection; c. positions of
low-energy Weyl points as well as d. energy band dispersions in the vicinity
of the Weyl point $k_{wp}=(0.46280,0.06931,0.02210)$. Point notations are as
follows: $%
K_{1x}=(0.36280,0.069310,0.022100),K_{2x}=(0.56280,0.069310,0.022100),K_{1y}=(0.46280,-0.17328,0.02210),K_{2y}=(0.46280,0.17328,0.022100),K_{1z}=(0.46280,0.06931,-0.11050),K_{2z}=(0.46280,0.06931,0.11050) 
$ in units $2\protect\pi /a,2\protect\pi /a,2\protect\pi /c$. Lattice
parameters used: a=12.1207 a.u., c/a=0.58513 \protect\cite{HfPRu}. }
\end{figure}
\end{center}

\begin{figure}[h]
\includegraphics[height=0.304\textwidth,width=0.95\textwidth]{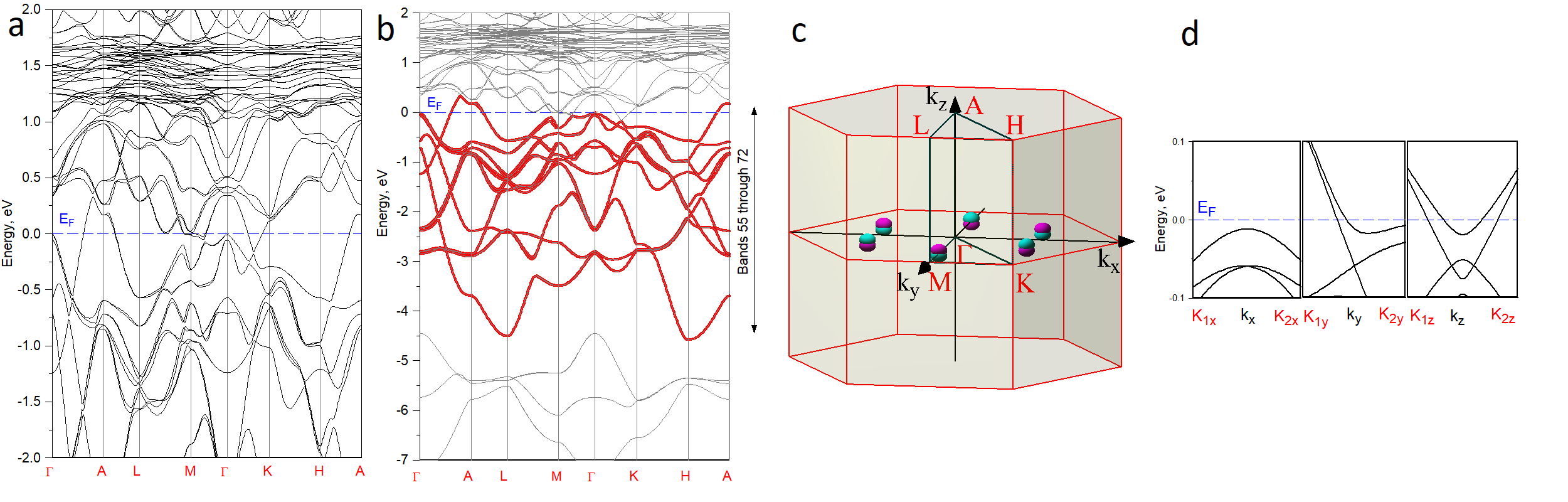}
\caption{Results for LaInMg: a. band structure near the Fermi level; b.
energy panel used for defining non-Abelian Berry connection; c. positions of
low-energy Weyl points as well as d. energy band dispersions in the vicinity
of the Weyl point $k_{wp}=(0.00000,0.36868,0.01123)$. Point notations are as
follows: $%
K_{1x}=(-0.10000,0.36868,0.01123),K_{2x}=(0.10000,0.36868,0.01123),K_{1y}=(0.0000,0.26868,0.01123),K_{2y}=(0.0000,0.46868,0.011230),K_{1z}=(0.0000,0.36868,-0.056150),K_{2z}=(0.0000,0.36868,0.056150) 
$ in units $2\protect\pi /a,2\protect\pi /a,2\protect\pi /c$. Lattice
parameters used: a=14.789 a.u., c/a=0.61472 \protect\cite{LaInMg}.}
\end{figure}

\begin{figure}[h]
\includegraphics[height=0.304\textwidth,width=0.95\textwidth]{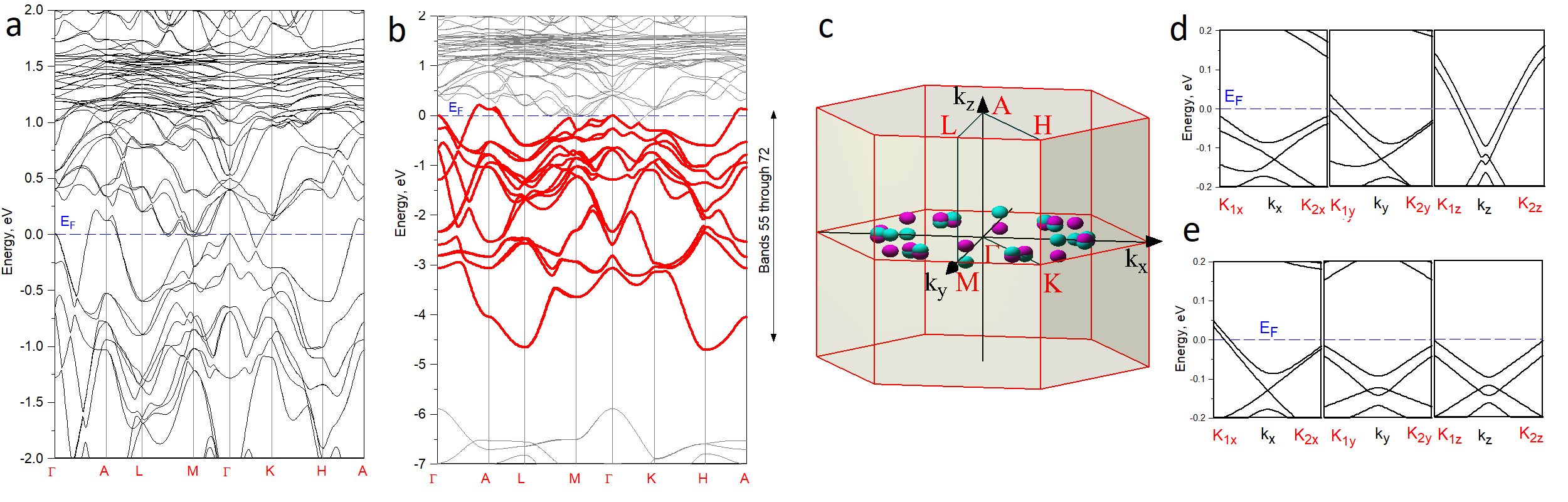}
\caption{Results for LaTlMg: a. band structure near the Fermi level; b.
energy panel used for defining non-Abelian Berry connection; c. positions of
low-energy Weyl points; d. energy band dispersions in the vicinity of the
Weyl point\ $k_{wp}=(0.00000,0.38916,0.03236)$. Points notations are as
follows: $%
K_{1x}=(-0.10000,0.38916,0.032360),K_{2x}=(0.10000,0.38916,0.032360),K_{1y}=(0.0000,0.28916,0.03236),K_{2y}=(0.0000,0.48916,0.03236),K_{1z}=(0.0000,0.38916,-0.16180),K_{2z}=(0.0000,0.38916,0.16180) 
$ in units $2\protect\pi /a,2\protect\pi /a,2\protect\pi /c$ as well as e.
energy band dispersions in the vicinity of the Weyl point\ $%
k_{wp}=(0.41450,0.02567,0.00724)$. Point notations are as follows: $%
K_{1x}=(0.3145,0,0.02567,0.00724),K_{2x}=(0.51450,0.02567,0.00724),K_{1y}=(0.41450,-0.12835,0.00724),K_{2y}=(0.41450,0.12835,0.00724),K_{1z}=(0.41450,0.02567,-0.0362),K_{2z}=(0.41450,0.02567,0.0362) 
$ in units $2\protect\pi /a,2\protect\pi /a,2\protect\pi /c$. Lattice
parameters used: a=14.7644 a.u., c/a=0.61160 \protect\cite{LaTlMg}.}
\end{figure}

\begin{figure}[h]
\includegraphics[height=0.304\textwidth,width=0.95\textwidth]{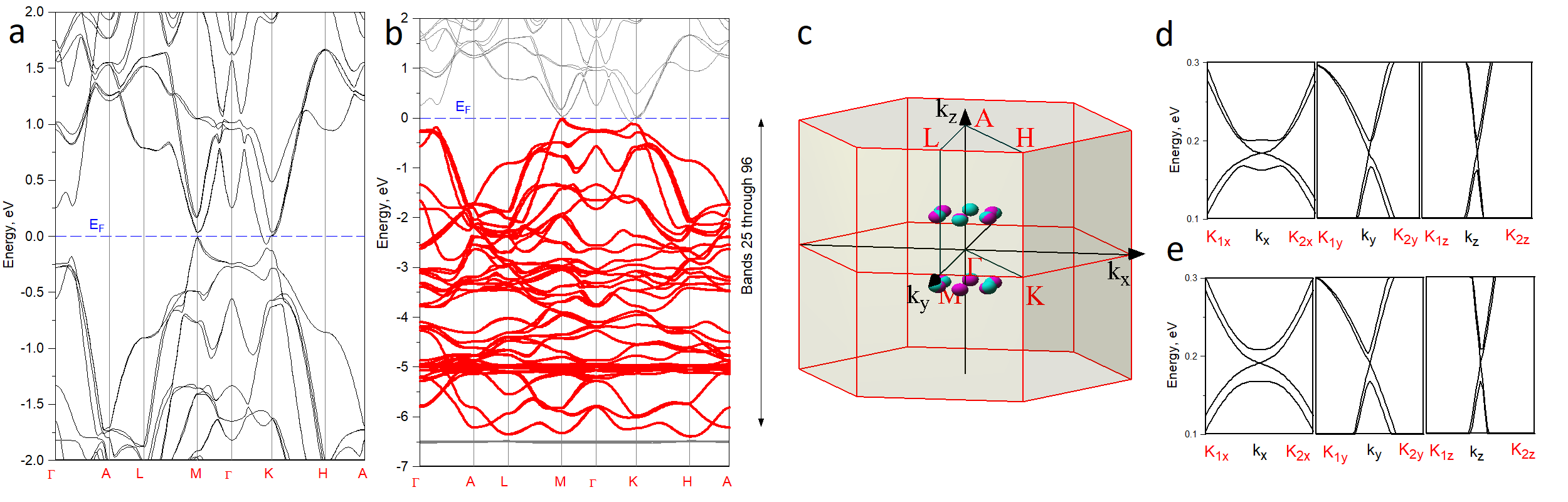}
\caption{Results for LuAsPd: a. band structure near the Fermi level; b.
energy panel used for defining non-Abelian Berry connection; c. positions of
low-energy Weyl points as well as d. energy band dispersions in the vicinity
of the Weyl point\ $k_{wp}=(0.00000,0.11481,0.14140)$. Point notations are
as follows: $%
K_{1x}=(-0.10000,0.11481,0.14140),K_{2x}=(0.10000,0.11481,0.14140),K_{1y}=(0.0000,0.01481,0.14140),K_{2y}=(0.0000,0.21481,0.14140),K_{1z}=(0.0000,0.11481,0.0414),K_{2z}=(0.0000,0.11481,0.24140) 
$ in units $2\protect\pi /a,2\protect\pi /a,2\protect\pi /c$. Lattice
parameters used: a=13.1733 a.u., c/a=0.55817 \protect\cite{LuAsPd}.}
\end{figure}

\begin{figure}[tbp]
\includegraphics[height=0.304\textwidth,width=0.95\textwidth]{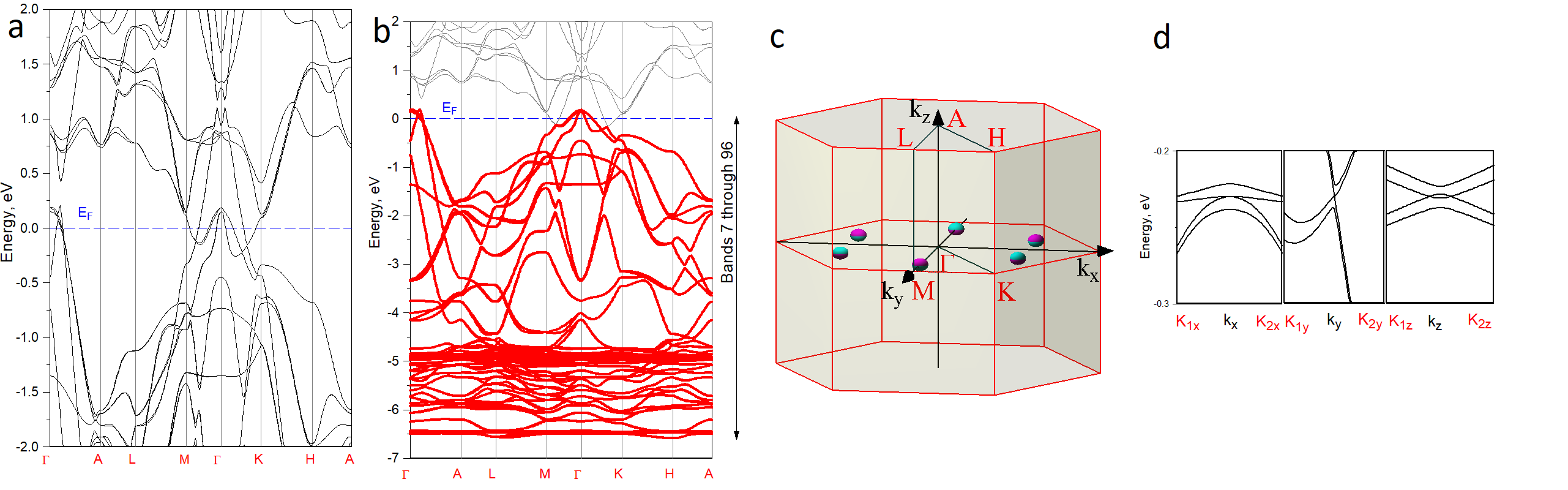}
\caption{Results for LuGeAg: a. band structure near the Fermi level; b.
energy panel used for defining non-Abelian Berry connection; c. positions of
low-energy Weyl points as well as d. energy band dispersions in the vicinity
of the Weyl point $k_{wp}=(0.00000,0.42190,0.00098)$. Point notations are as
follows: $%
K_{1x}=(-0.10000,0.42190,0.00098),K_{2x}=(0.10000,0.42190,0.00098),K_{1y}=(0.0000,0.32190,0.00098),K_{2y}=(0.0000,0.52190,0.00098),K_{1z}=(0.0000,0.42190,-0.0049),K_{2z}=(0.0000,0.42190,0.0049) 
$ in units $2\protect\pi /a,2\protect\pi /a,2\protect\pi /c$. Lattice
parameters used: a=13.2517 a.u., c/a=0.58948 \protect\cite{LuGeAg}.}
\end{figure}

\begin{figure}[tbp]
\includegraphics[height=0.304\textwidth,width=0.95\textwidth]{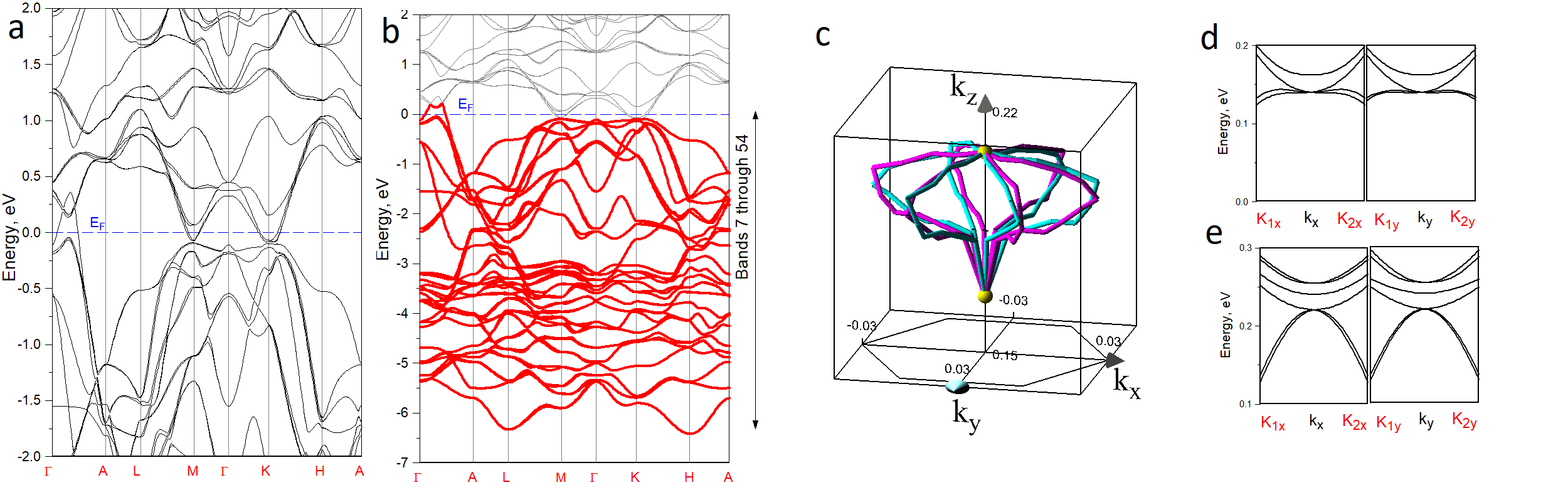}
\caption{Results for TiGePd: a. band structure near the Fermi level; b.
energy panel used for defining non-Abelian Berry connection; c. nodal lines
and positions of triple degenerate points. The zoomed area of the BZ is
bounded by $0.15\leq 2\protect\pi k_{z}/c\leq 0.22$ and $-0.03\leq 2\protect%
\pi k_{x,y}/a\leq +0.03;$ d. energy band dispersions in the vicinity of the
triple point\ $k_{tp}=(0.00000,0.00000,0.16495)$. Points notations are as
follows: $%
K_{1x}=(-0.10000,0.0000,0.16495),K_{2x}=(0.10000,0.0000,0.16495),K_{1y}=(0.0000,-0.10000,0.16495),K_{2y}=(0.0000,0.10000,0.16495) 
$ in units $2\protect\pi /a,2\protect\pi /a,2\protect\pi /c$; e. energy band
dispersions in the vicinity of the triple point\ $%
k_{tp}=(0.00000,0.00000,0.20775)$. Point notations are as follows: $%
K_{1x}=(-0.10000,0.0000,0.20775),K_{2x}=(0.10000,0.0000,0.20775),K_{1y}=(0.0000,-0.10000,0.20775) 
$ in units $2\protect\pi /a,2\protect\pi /a,2\protect\pi /c$. Lattice
parameters used: a=12.4779 a.u., c/a=0.56032 \protect\cite{TiGePd}.}
\end{figure}

\begin{figure}[tbp]
\includegraphics[height=0.304\textwidth,width=0.95\textwidth]{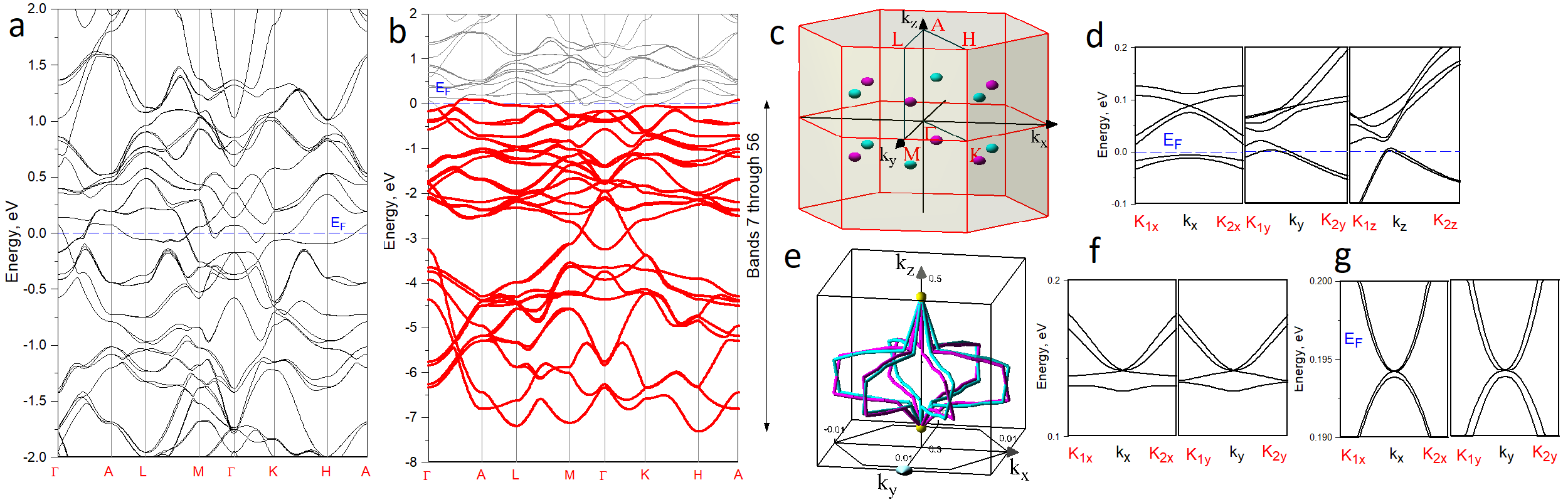}
\caption{Results for VAsFe: a. band structure near the Fermi level; b.
energy panel used for defining non-Abelian Berry connection; c. positions of
low-energy Weyl points; d. energy band dispersions in the vicinity of the
Weyl point\ $k_{wp}=(0.00000,0.38339,0.17269).$ Points notations are as
follows: $%
K_{1x}=(-0.10000,0.38339,0.17269),K_{2x}=(0.10000,0.38339,0.17269),K_{1y}=(0.0000,0.28339,0.17269),K_{2y}=(0.0000,0.48339,0.17269),K_{1z}=(0.0000,0.38339,0.07269),K_{2z}=(0.0000,0.38339,0.27269) 
$ in units $2\protect\pi /a,2\protect\pi /a,2\protect\pi /c$.; e. nodal
lines with triple degenerate points. The zoomed area of the BZ is bounded by 
$0.3\leq 2\protect\pi k_{z}/c\leq 0.5$ and $-0.01\leq 2\protect\pi %
k_{x,y}/a\leq +0.01.$; f. energy band dispersions in the vicinity of the
triple point\ $k_{tp}=(0.00000,0.00000,0.32279)$. Points notations are as
follows: $%
K_{1x}=(-0.10000,0.0000,0.32279),K_{2x}=(0.10000,0.0000,0.32279),K_{1y}=(0.0000,-0.1000,0.32279),K_{2y}=(0.00000,0.1000,0.32279). 
$ in units $2\protect\pi /a,2\protect\pi /a,2\protect\pi /c$; g. energy band
dispersions in the vicinity of the triple point\ $%
k_{tp}=(0.00000,0.00000,0.47625)$. Point notations are as follows: $%
K_{1x}=(-0.10000,0.0000,0.47625),K_{2x}=(0.10000,0.0000,0.47625),K_{1y}=(0.0000,-0.1000,0.47625),K_{2y}=(0.00000,0.1000,0.47625). 
$ in units $2\protect\pi /a,2\protect\pi /a,2\protect\pi /c$. Lattice
parameters used: a=11.7352 a.u., c/a=0.56892 \protect\cite{VAsFe}.}
\end{figure}

\begin{figure}[tbp]
\includegraphics[height=0.304\textwidth,width=0.95\textwidth]{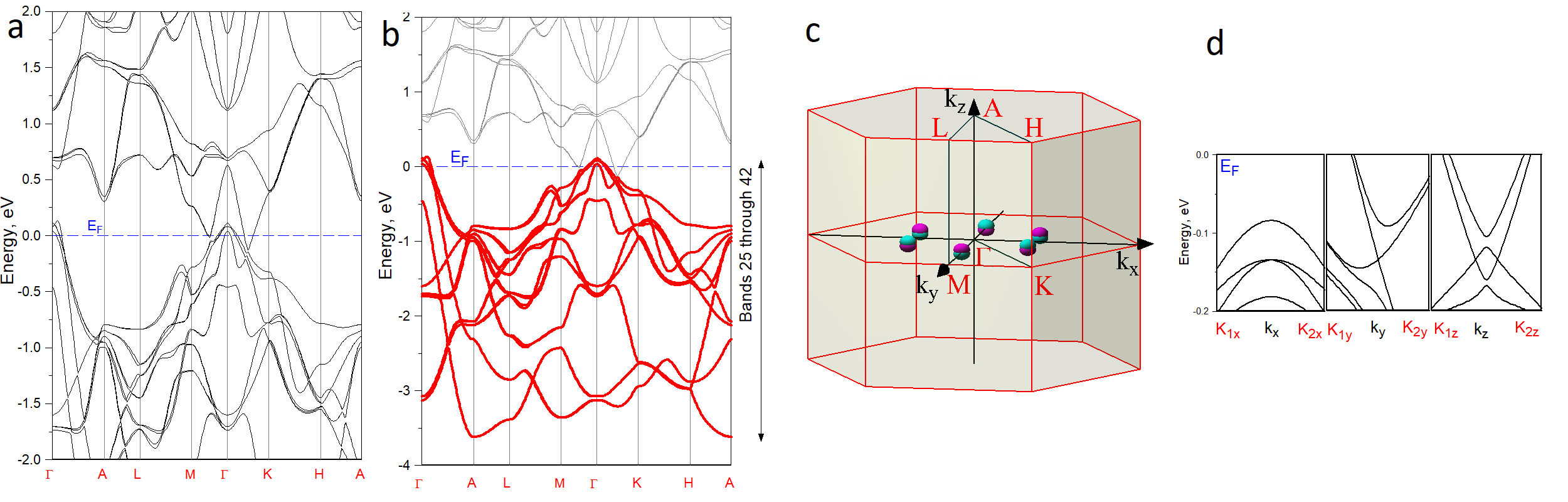}
\caption{Results for YGeLi: a. band structure near the Fermi level; b.
energy panel used for defining non-Abelian Berry connection; c. positions of
low-energy Weyl points as well as d. energy band dispersions in the vicinity
of the Weyl point $k_{wp}=(0.00000,0.27793,0.00817)$. Point notations are as
follows: $%
K_{1x}=(-0.10000,0.27793,0.00817),K_{2x}=(0.10000,0.27793,0.00817),K_{1y}=(0.0000,0.17793,0.00817),K_{2y}=(0.0000,0.37793,0.00817),K_{1z}=(0.0000,0.27793,-0.040850),K_{2z}=(0.0000,0.27793,0.04085) 
$ in units $2\protect\pi /a,2\protect\pi /a,2\protect\pi /c$. Lattice
parameters used: a=13.3509 a.u., c/a=0.59915 \protect\cite{YGeLi}.}
\end{figure}

\begin{figure}[tbp]
\includegraphics[height=0.304\textwidth,width=0.95\textwidth]{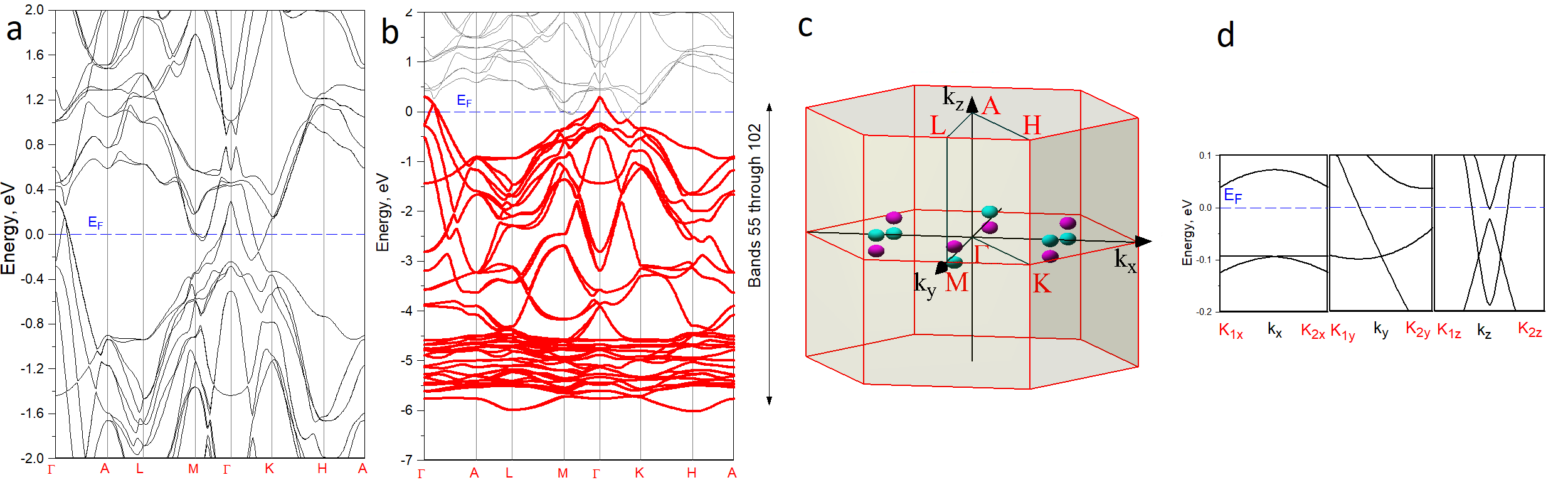}
\caption{Results for YPbAg: a. band structure near the Fermi level; b.
energy panel used for defining non-Abelian Berry connection; c. positions of
low-energy Weyl points as well as d. energy band dispersions in the vicinity
of the Weyl point $k_{wp}=(0.00000,0.40335,0.03142)$. Point notations are as
follows: $%
K_{1x}=(-0.10000,0.40335,0.03142),K_{2x}=(0.10000,0.40335,0.03142),K_{1y}=(0.0000,0.30335,0.03142),K_{2y}=(0.0000,0.50335,0.03142),K_{1z}=(0.0000,0.40335,-0.15710),K_{2z}=(0.0000,0.40335,0.15710) 
$ in units $2\protect\pi /a,2\protect\pi /a,2\protect\pi /c$. Lattice
parameters used: a=14.140 a.u., c/a=0.59133 \protect\cite{YPbAg}.}
\end{figure}

\begin{figure}[tbp]
\includegraphics[height=0.304\textwidth,width=0.95\textwidth]{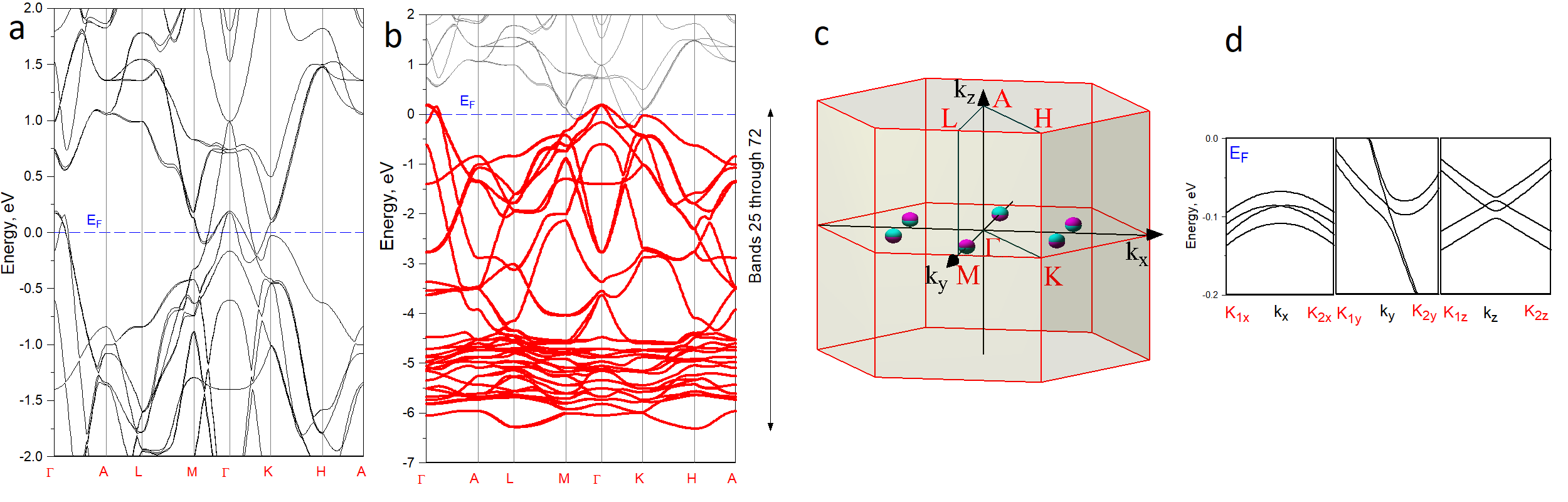}
\caption{Results for YSiAg: a. band structure near the Fermi level; b.
energy panel used for defining non-Abelian Berry connection; c. positions of
low-energy Weyl points as well as d. energy band dispersions in the vicinity
of the Weyl point $k_{wp}=(0.00000,0.37866,0.00385)$. Point notations are as
follows: $%
K_{1x}=(-0.10000,0.37864,0.00385),K_{2x}=(0.10000,0.37864,0.00385),K_{1y}=(0.0000,0.27864,0.00385),K_{2y}=(0.0000,0.47864,0.00385),K_{1z}=(0.0000,0.37864,-0.0192),K_{2z}=(0.0000,0.37864,0.0192) 
$ in units $2\protect\pi /a,2\protect\pi /a,2\protect\pi /c$. Lattice
parameters used: a=13.2623 a.u., c/a=0.59364 \protect\cite{YSiAg}.}
\end{figure}

\begin{figure}[tbp]
\includegraphics[height=0.304\textwidth,width=0.95\textwidth]{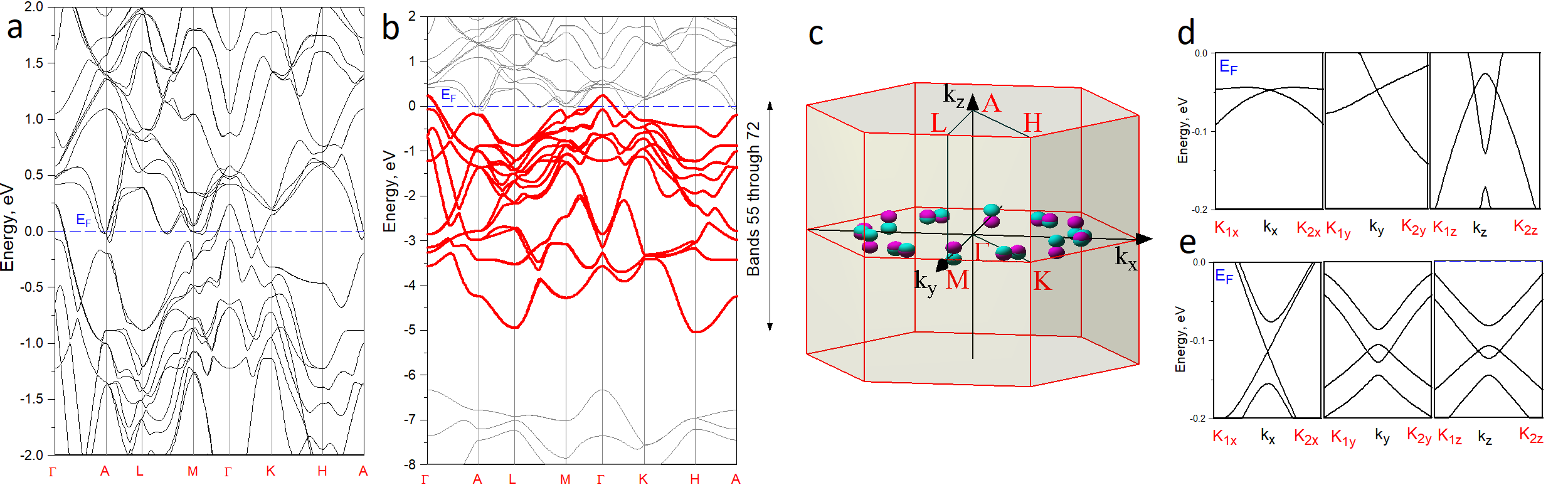}
\caption{Results for YTlMg: a. band structure near the Fermi level; b.
energy panel used for defining non-Abelian Berry connection; c. positions of
low-energy Weyl points; d. energy band dispersions in the vicinity of the
Weyl point $k_{wp}=(0.00000,0.43303,0.02319)$. Points notations are as
follows: $%
K_{1x}=(-0.10000,0.43303,0.02319),K_{2x}=(0.10000,0.43303,0.02319),K_{1y}=(0.0000,0.33303,0.02319),K_{2y}=(0.0000,0.53303,0.02319),K_{1z}=(0.0000,0.43303,-0.11595),K_{2z}=(0.0000,0.43303,0.11595) 
$ in units $2\protect\pi /a,2\protect\pi /a,2\protect\pi /c$. e. energy band
dispersions in the vicinity of the Weyl point $%
k_{wp}=(0.44076,0.02908,0.00441)$. Point notations are as follows: $%
K_{1x}=(0.34076,0.02908,0.00441),K_{2x}=(0.54076,0.02908,0.00441),K_{1y}=(0.44076,-0.14540,0.00441),K_{2y}=(0.44076,0.14540,0.00441),K_{1z}=(0.44076,0.02908,-0.02205),K_{2z}=(0.44076,0.02908,0.02205) 
$ in units $2\protect\pi /a,2\protect\pi /a,2\protect\pi /c$. Lattice
parameters used: a=14.1824 a.u., c/a=0.61272 \protect\cite{LaTlMg}.}
\end{figure}

\begin{figure}[tbp]
\includegraphics[height=0.304\textwidth,width=0.95\textwidth]{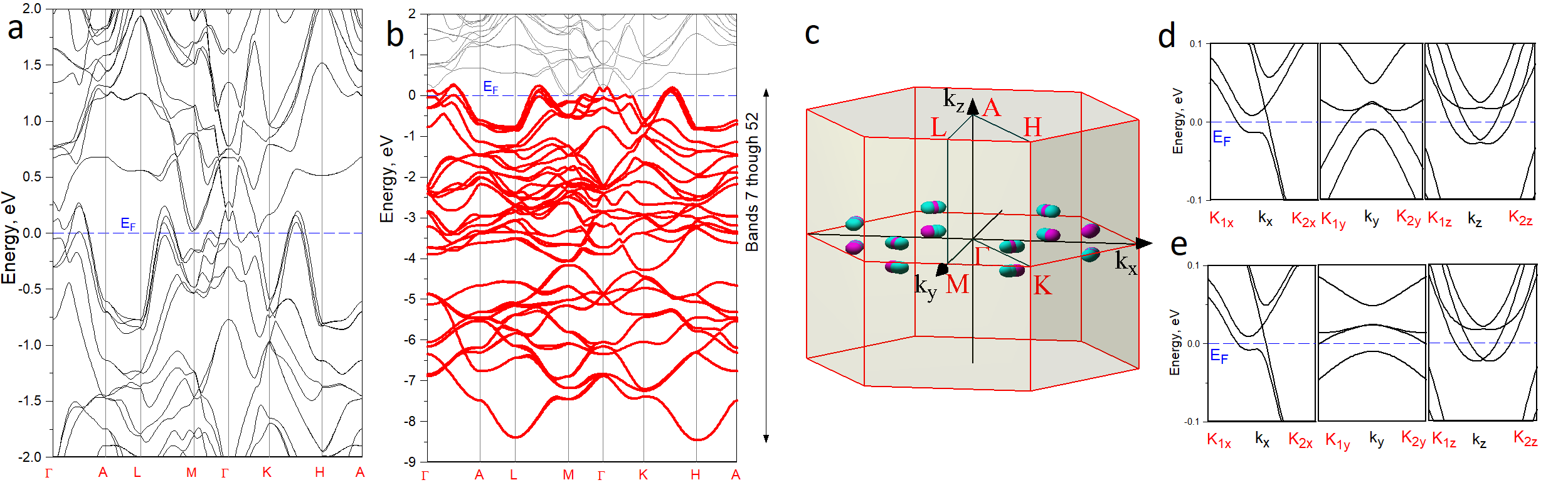}
\caption{Results for ZrAsOs: a. band structure near the Fermi level; b.
energy panel used for defining non-Abelian Berry connection; c. positions of
low-energy Weyl points; d. energy band dispersions in the vicinity of the
Weyl point $k_{wp}=(0.47365,0.02591,0.04792)$. Points notations are as
follows: $%
K_{1x}=(0.37365,0.02591,0.04792),K_{2x}=(0.57365,0.02591,0.04792),K_{1y}=(0.47365,-0.12955,0.04792),K_{2y}=(0.47365,0.12955,0.04792),K_{1z}=(0.47365,0.02591,-0.11980),K_{2z}=(0.47365,0.02591,0.11980) 
$ in units $2\protect\pi /a,2\protect\pi /a,2\protect\pi /c$; e. energy band
dispersions in the vicinity of the Weyl point $%
k_{wp}=(0.474060.012150.047890)$. Point notations are as follows: $%
K_{1x}=(0.37406,-0.01215,0.04789),K_{2x}=(0.57406,-0.01215,0.04789),K_{1y}=(0.47406,-0.06075,0.04789),K_{2y}=(0.47406,0.06075,0.04789),K_{1z}=(0.47406,0.01215,-0.11973),K_{2z}=(0.47406,0.01215,0.11973) 
$ in units $2\protect\pi /a,2\protect\pi /a,2\protect\pi /c$. Lattice
parameters used: a=12.476 a.u., c/a=0.57467 \protect\cite{ZrAsOs}.}
\end{figure}
\begin{figure}[h]
\includegraphics[height=0.304\textwidth,width=0.95\textwidth]{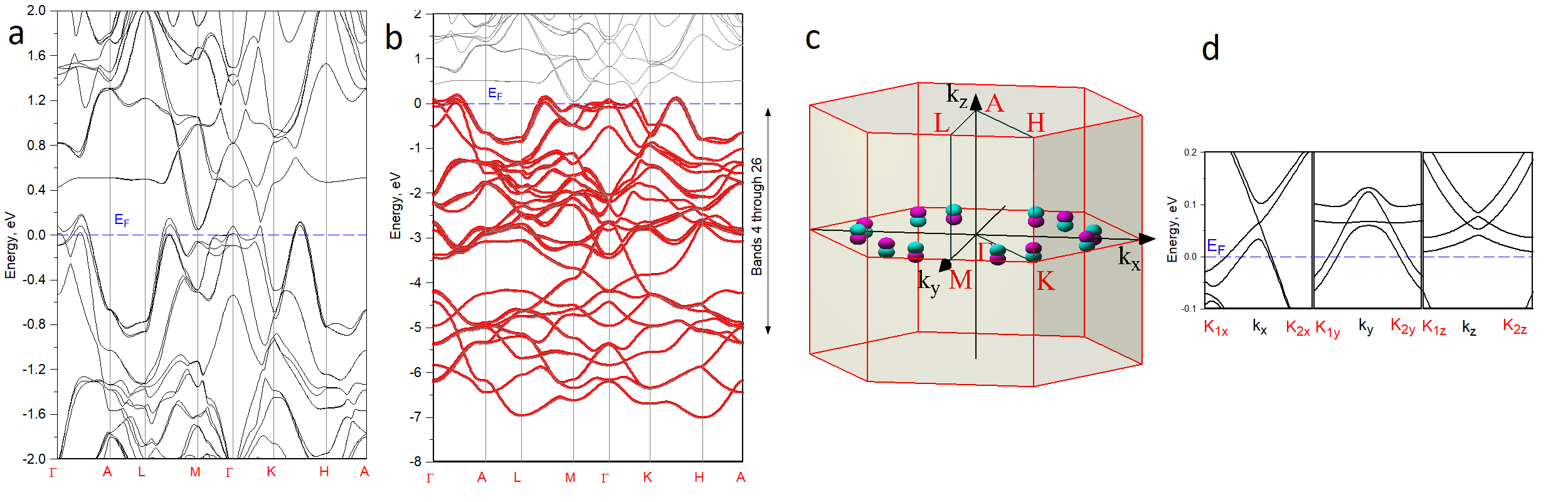}
\caption{Results for ZrPRu: a. band structure near the Fermi level; b.
energy panel used for defining non-Abelian Berry connection; c. positions of
low-energy Weyl points as well as d. energy band dispersions in the vicinity
of the Weyl point\ $k_{wp}=(0.45982,0.07532,0.01698)$. Point notations are
as follows: $%
K_{1x}=(0.35982,0.07532,0.01698),K_{2x}=(0.55982,0.07532,0.01698),K_{1y}=(0.45982,-0.18830,0.01698),K_{2y}=(0.45982,0.18830,0.01698),K_{1z}=(0.45982,0.07532,-0.0849),K_{2z}=(0.45982,0.07532,0.0849) 
$ in units $2\protect\pi /a,2\protect\pi /a,2\protect\pi /c$. Lattice
parameters used: a=12.2057 a.u., c/a=0.58492 \protect\cite{HfPRu}.}
\end{figure}

\end{widetext}